\documentclass[traditabstract]{aa} 
\usepackage{graphicx}
\usepackage{txfonts}
\usepackage{url}
\usepackage{natbib}
\bibpunct{(}{)}{;}{a}{}{,} 
\usepackage[utf8]{inputenc}
\usepackage{subfigure}
\usepackage{verbatim} 
\begin{document}
   \title{MHD wave transmission in the Sun's atmosphere}
   \author{M. Stangalini$^{1}$, D. Del Moro$^{1}$, F. Berrilli$^{1}$, S. M. Jefferies$^{2}$}
   \institute{$^{1}$Dipartimento di Fisica, Università di Roma "Tor Vergata"\\
   Via della Ricerca Scientifica 1, 00133 Rome, Italy\\
   			  $^{2}$University of Hawai'i, Institute for Astronomy\\
   34 `Ohi`a Ku St. Pukalani, Hawaii 96768-8288, USA \\
              \email{marco.stangalini@roma2.infn.it}}
             
   \date{Draft version}

 
  \abstract
   {MHD wave propagation inside the Sun's atmosphere is closely related to the magnetic field topology. For example, magnetic fields are able to lower the cutoff frequency for acoustic waves thus allowing waves which would otherwise be trapped below the photosphere to propagate into the upper atmosphere. Another example is that MHD waves can be transmitted or converted into other forms of waves at altitudes where the sound speed equals the Alfv\'en speed. We take advantage of the large field-of-view provided by the IBIS experiment to study the wave propagation at two heights in the solar atmosphere, as sampled by the photospheric Fe $617.3$ nm spectral line and the chromospheric Ca $854.2$ nm spectral line, and its relationship to the local magnetic field. Among other things, we find substantial leakage of waves with 5-minute periods in the chromosphere at the edges of a pore and in the diffused magnetic field surrounding it. By using spectro-polarimetric inversions of Hinode SOT/SP data, we also find a relationship between the photospheric power spectrum and the magnetic field inclination angle; in particular, well-defined transmission peaks around $25\degr$ for $5$ minutes waves and around $15\degr$ for $3$ minutes waves. We propose a very simple model based upon wave transmission theory to explain this behavior. Finally, an analysis of both the power spectra and chromospheric amplification spectra suggests the presence of longitudinal acoustic waves along the magnetic field lines.}

   \authorrunning{M. Stangalini}

\maketitle

%

\section{Introduction}
The source of the energy needed to heat the Sun's corona is a long-standing enigma. The two main competing theories proposed to explain the energy excess in the outer atmosphere of the Sun are Joule heating through resistive dissipation by magnetic field reconnection and mechanical heating by waves \citep[see][for a review of the mechanisms proposed]{2001ApJ...560.1035A}. However, both Joule heating and mechanical heating by high-frequency acoustic waves ($\nu_{c}>5.2$ mHz) have recently been ruled out as major contributors to the energy budget of the corona \citep{2006ApJ...646..579F,npg-16-443-2009}.
Low-frequency MHD waves ($\nu < 5$ mHz), on the other hand, may represent a significant source of energy \citep{2006ApJ...648L.151J}. Though they are generally not allowed to propagate into the atmosphere, as their frequency does not exceed the expected photospheric cutoff frequency ($\nu_{c}=5.2$ mHz), in regions where the magnetic field is largely inclined with respect to the gravity vector the cutoff frequency can be substantially lowered through the ramp effect:
\begin{equation}
\omega_{eff}=\omega_{c} ~cos~ \theta
\end{equation}
where $\omega_{eff}$ is the effective cutoff frequency and $\theta$ is the angle of the magnetic field to the local gravity vector. This effect is behind the so-called magneto-acoustic portals (\cite{2006ApJ...648L.151J}) which allow waves with frequencies far below $5.2$ mHz to be channeled into the strongly inclined magnetic field.\\
It has been demonstrated, both theoretically \citep{2006MNRAS.372..551S,2008SoPh..251..251C} and through numerical simulations \citep{2006ApJ...653..739K, 2010ApJ...719..357F}, that at locations where the sound speed $c_{s}$ and Alfv\'en speed $a$ nearly coincide, part of the energy contained in the acoustic-like component (fast MHD mode in the $\beta>1$ regime) can be converted to two types of waves: field aligned acoustic waves (slow MHD mode in $\beta<1$ plasma), or magnetic-like waves (fast mode in $\beta<1$ regions). These two processes are commonly referred to as "fast to slow" and "fast to fast" conversion. In the first case, the acoustic nature of the wave is preserved while, in the fast to fast conversion, the wave changes from acoustic-like to magnetic-like. We note that since $c^{2}_{s}/a^{2}=(\gamma/2) \beta$, the layer where the gas pressure is equal to the magnetic pressure (the $\beta=1$ or equipartition layer) is in practice very close to the layer where the phase speed of the fast and slow modes coincide, even if they are conceptually different. The amount of energy transferred to the acoustic-like mode or converted into the magnetic-like mode, as the wave crosses the equipartition layer, depends on the angle between the wavevector and the magnetic field (the attack angle $\alpha$):

    \begin{figure*}[!]
   \centering
   \subfigure[Core Fe 617.3 nm]{\includegraphics		[width=8cm,trim=12mm 3mm 12mm 3mm, clip]{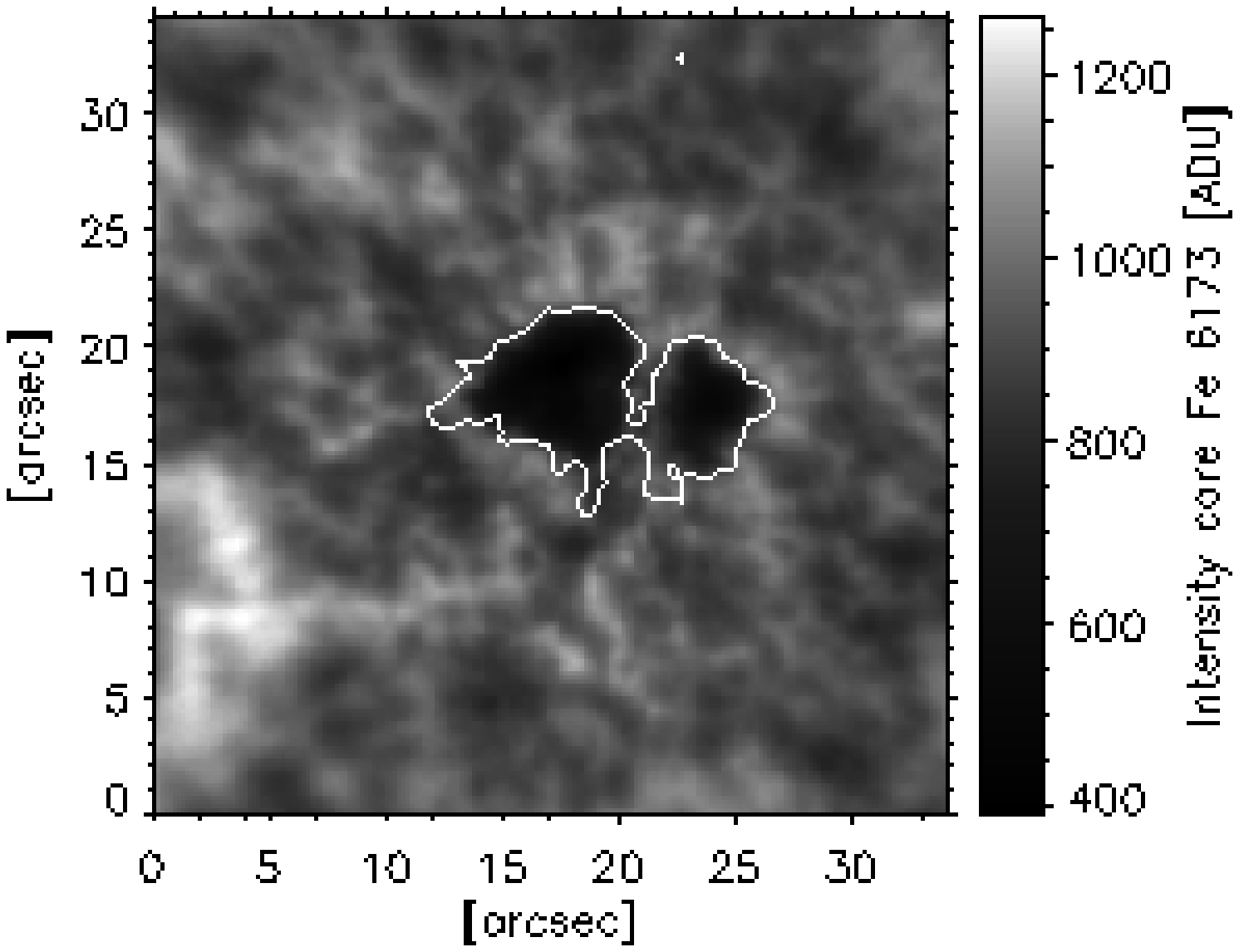}}
   \subfigure[Core Ca 854.2 nm]{\includegraphics		[width=8cm,trim=12mm 3mm 12mm 3mm, clip]{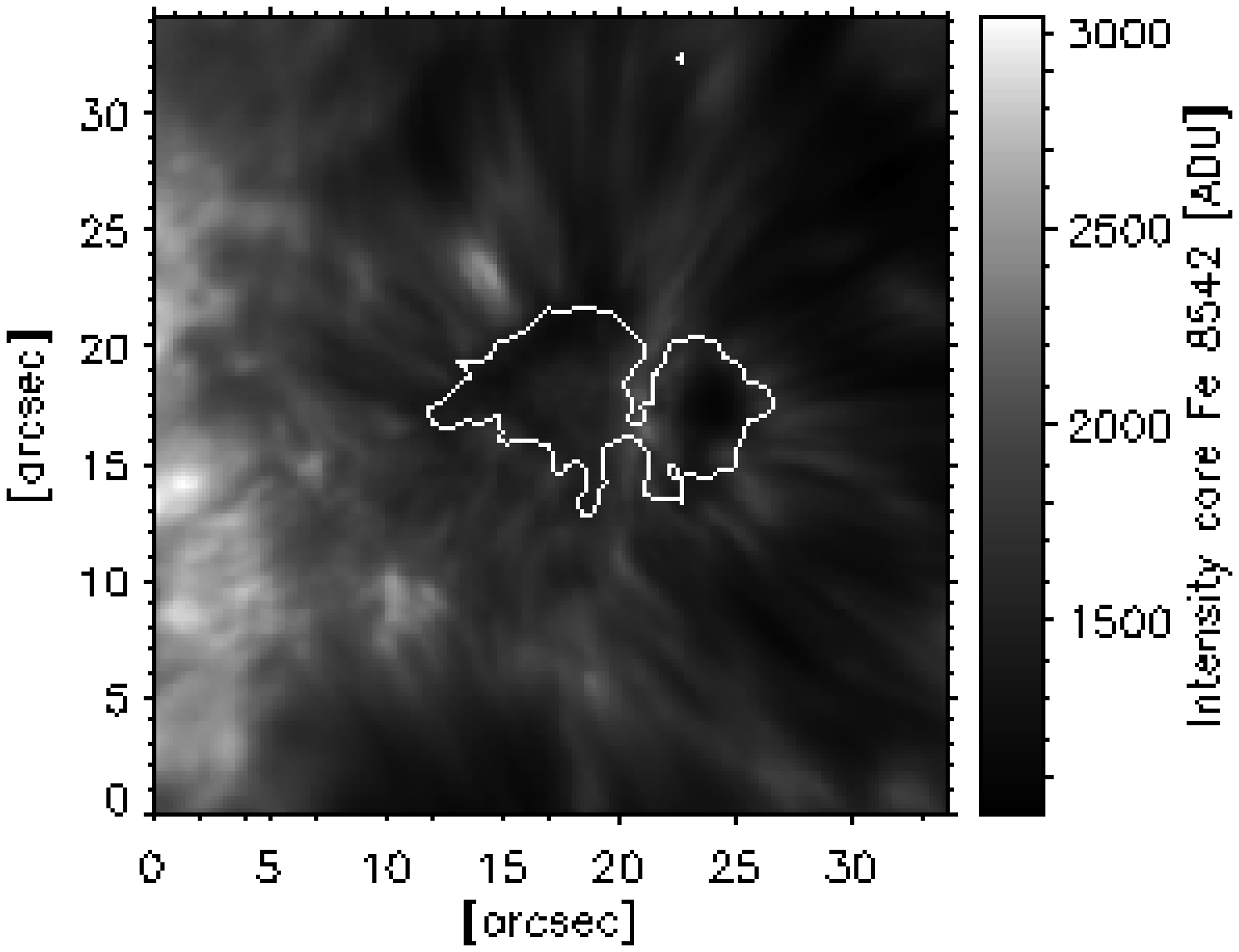}}
   \subfigure[$\theta$ @ $300$ km]{\includegraphics		[width=8cm,trim=12mm 3mm 12mm 3mm, clip]{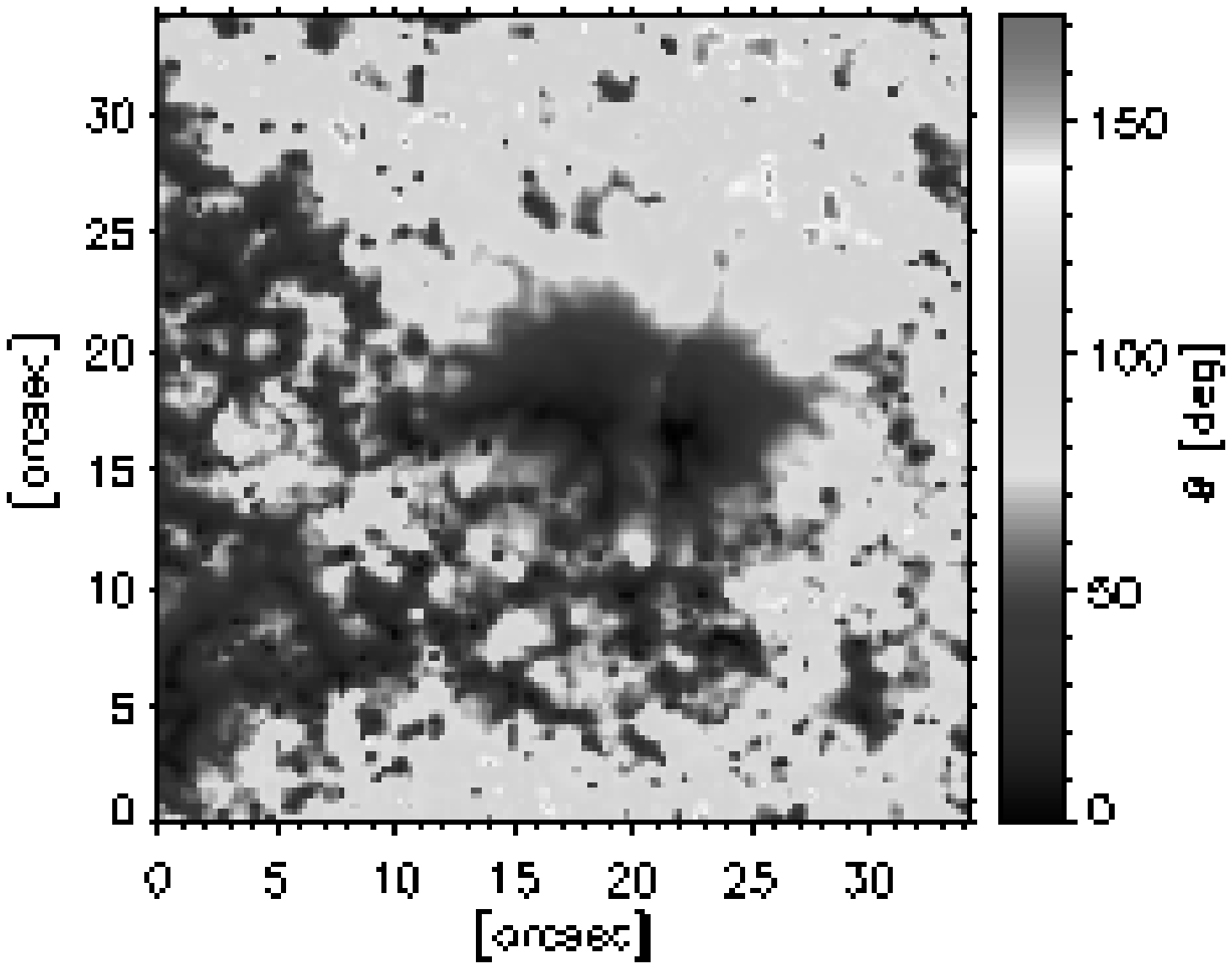}}
    \caption{(a) IBIS intensity image in the core of the Fe $617.3$ nm photospheric line. (b)IBIS intensity image in the core of the Ca $854.2$ nm chromospheric line. (c) Magnetic field inclination in the photosphere as inferred from MERLIN inversions on Hinode data.} 
    \label{inclination}
   \end{figure*}

\begin{equation}
T=exp[-\pi k h_{s} sin^{2}\alpha]
\end{equation}
where $h_{s}$ is the thickness of the equipartition layer as measured along the direction of propagation, in which the process of mode conversion is taking place, $k$ the wavenumber and $T$ the "fast-to-slow" transmission coefficient. It is important to note that this relation is strictly valid only for small attack angles, a comparison with the exact solution can be found in \cite{2009SoPh..255..193H}. The "fast-to-fast" transmission coefficient $C$ is obtained by invoking conservation of energy, i.e.
\begin{equation}
T+|C|=1
\end{equation}
and $C$ is complex (as we need to take into account possible phase changes during conversion).
\citet{2006MNRAS.372..551S} demonstrated that the ramp and mode conversion effects together result in a strong dependence of the acoustic energy flux on both the magnetic field inclination and the attack angle. In particular, the acoustic flux should have a maximum for magnetic field inclination angles between $20$ and $30$ degrees. This is a result of the transmission coefficient being large at smaller attack angles and the ramp effect allowing the propagation of low frequency waves once $cos~\theta < \omega_{eff}/ \omega_{c}$, that is at large inclination angles.
Here we investigate this claim and show how the velocity field fluctuations of a solar active region observed by the Interferometric BIdimensional Spectrometer (IBIS) based on a dual Fabry-Perot system, depend on the inclination angle of the magnetic field, as inferred from the spectropolarimetric inversions of the same region observed by Hinode SOT/SP \citep{springerlink:10.1007/s11207-008-9174-z}.\\
Among other things, we find that the power of the velocity oscillations is dependent on both the frequency and on the magnetic field inclination: this is in accord with the above theoretical picture. This scenario is also supported by the analysis of the spatial distribution of the power in the chromosphere, which shows that there is a substantial lack of power at the locations where the magnetic field is bent horizontally. This is consistent with longitudinal waves moving along the field lines (which can't be detected when they are propagating parallel to the line-of-sight).  

\begin{figure*}
\centering
\subfigure[Fe 6173 @ $3$ mHz]{\includegraphics[width=6cm,trim=12mm 3mm 12mm 3mm, clip]{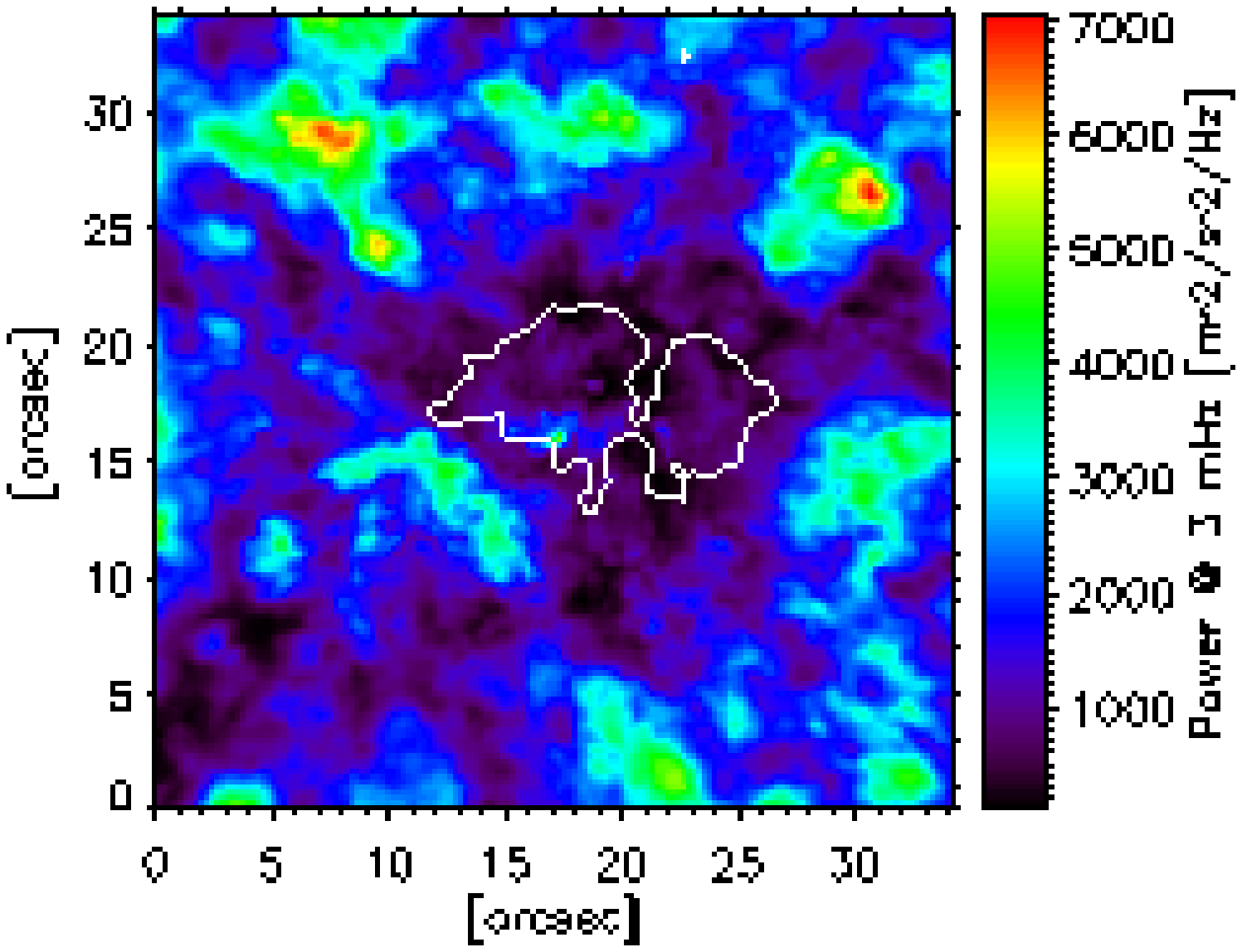}}
\subfigure[Ca 8542 @ $3$ mHz]{\includegraphics[width=6cm,trim=12mm 3mm 12mm 3mm, clip]{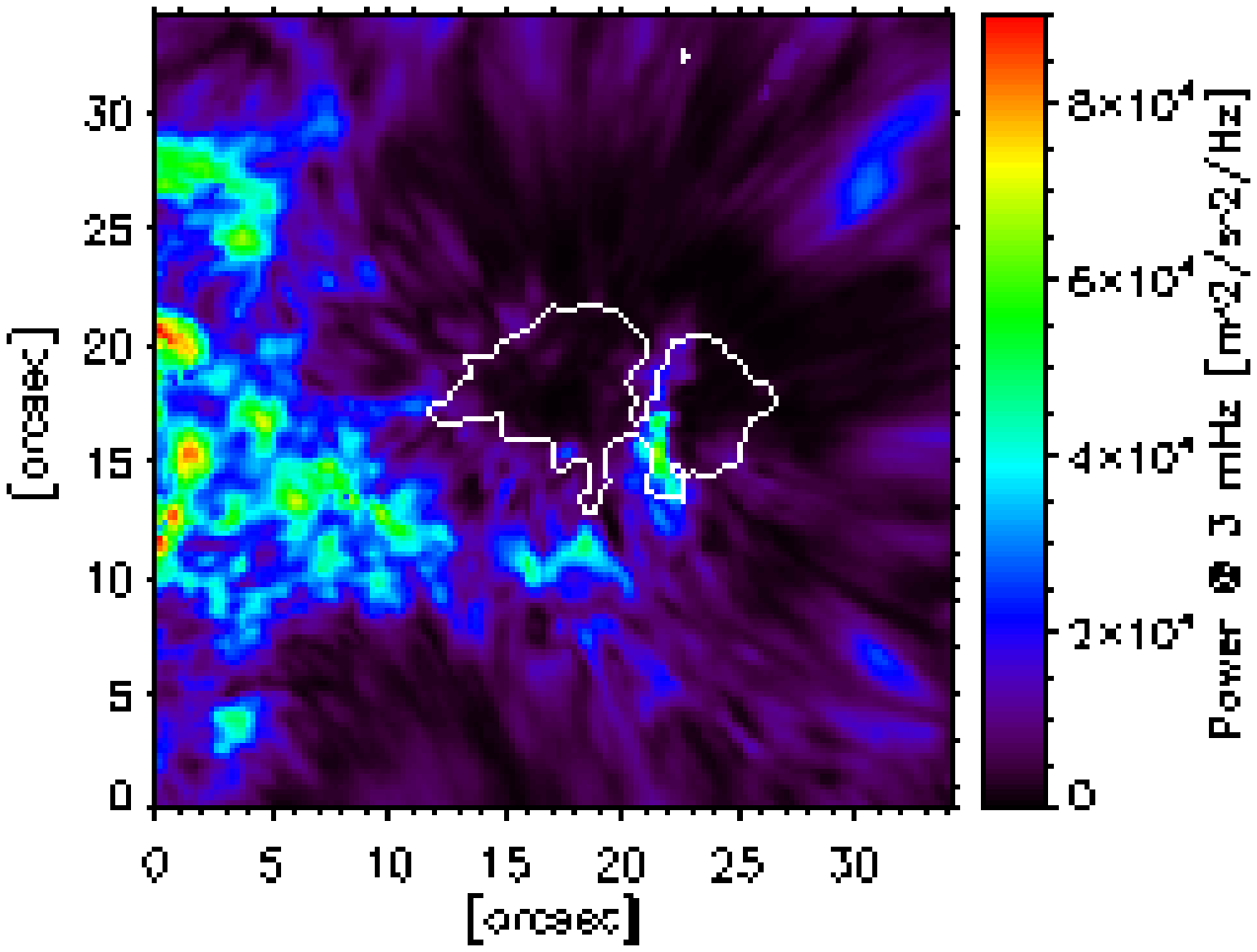}}
\subfigure[Fe 6173 @ $5$ mHz]{\includegraphics[width=6cm,trim=12mm 3mm 12mm 3mm, clip]{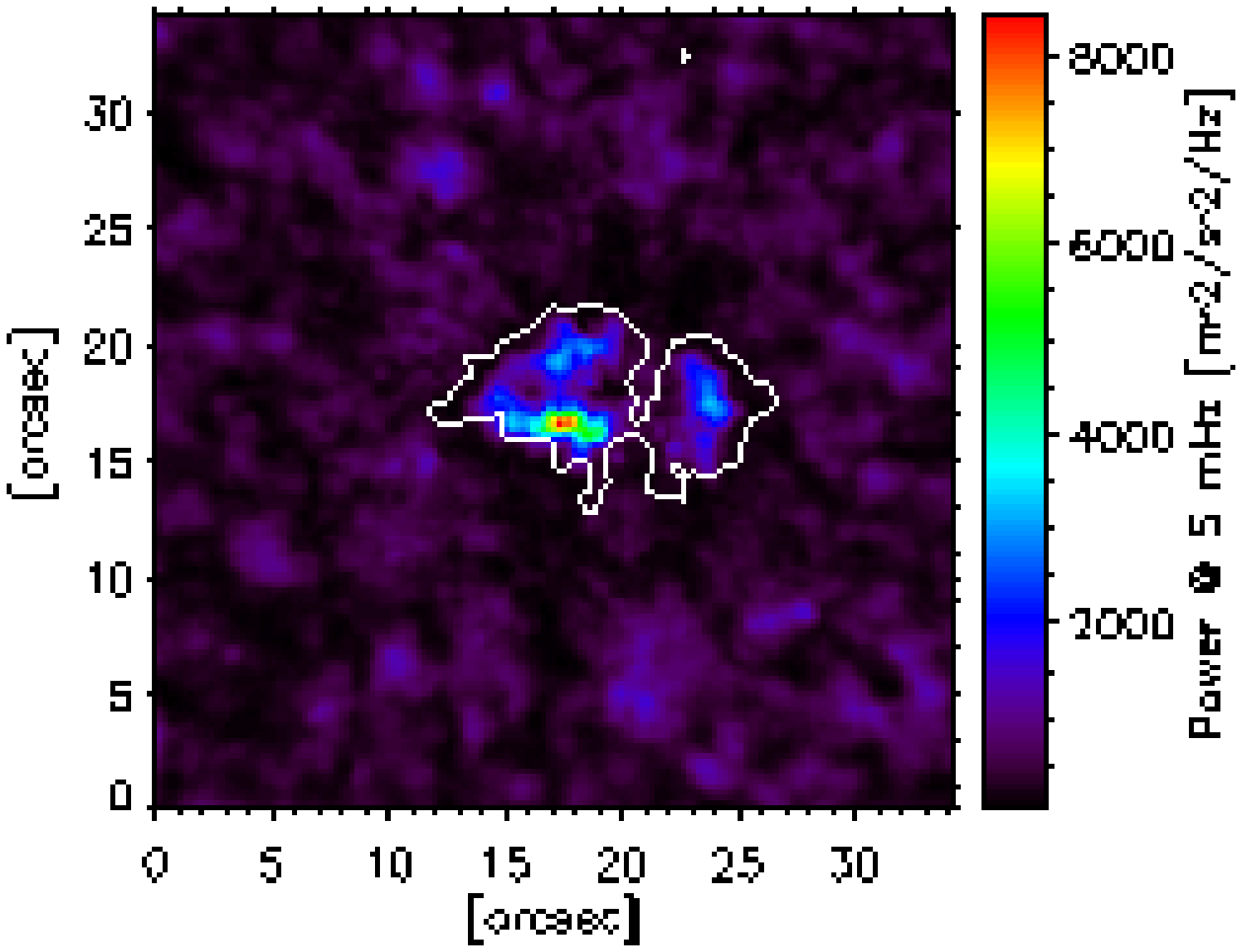}}
\subfigure[Ca 8542 @ $5$ mHz]{\includegraphics[width=6cm,trim=12mm 3mm 12mm 3mm, clip]{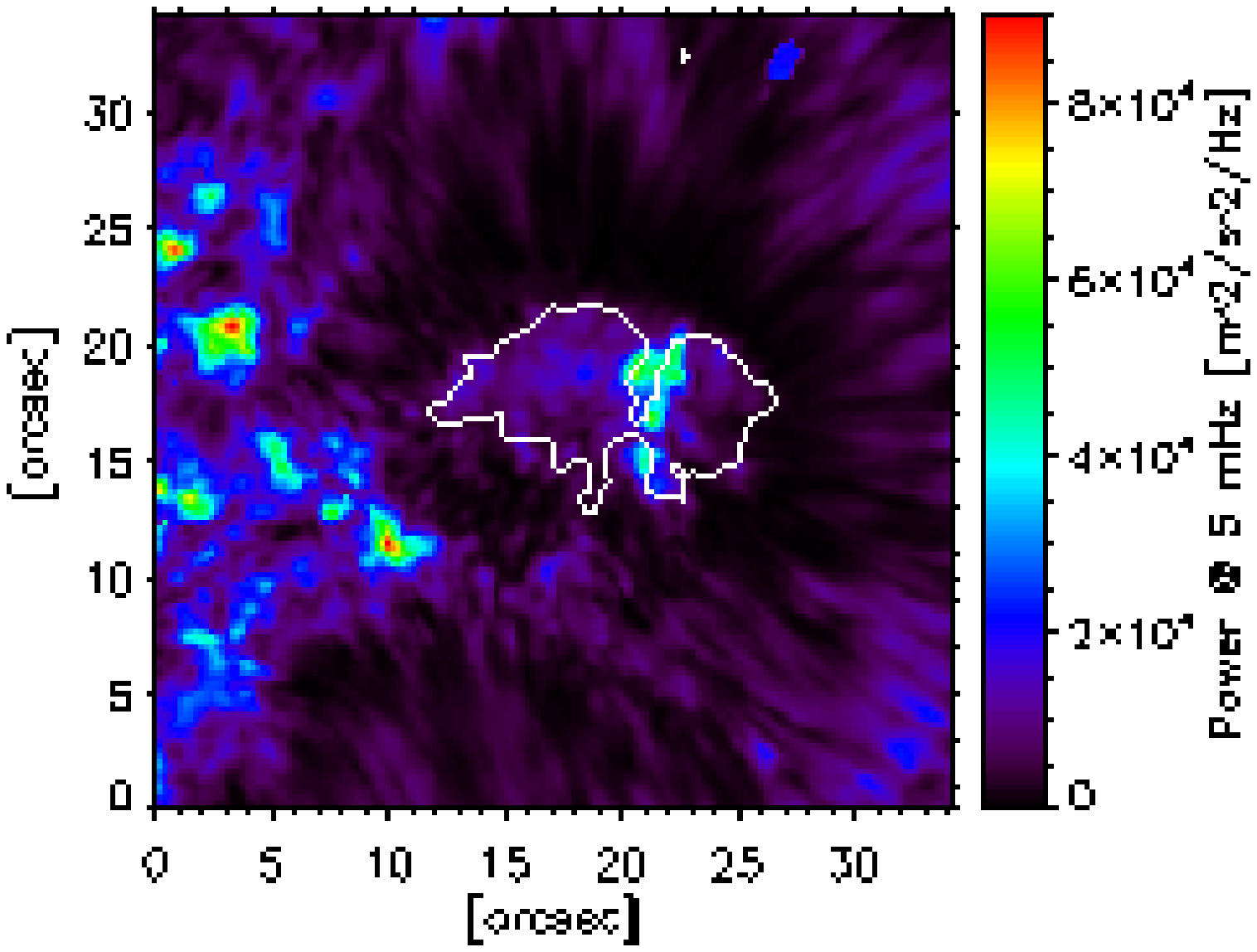}}

\caption{(a) IBIS $Fe~617.3$ nm power map at $3~mHz$. (b) IBIS $Ca~854.2$ nm power map at $3~mHz$. (c) IBIS $Fe~617.3$ nm power map at $5~mHz$. (d) IBIS $Ca~854.2$ nm power map at $5~mHz$. Continuous contours indicate approximatively the position of the umbra.}
\label{power}
\end{figure*}

\section{Observations}  
The data used in the work were acquired on October 15th 2008 at $4:30$ pm UT in full Stokes mode with IBIS at the Dunn Solar Telescope.  The experiment combined high-spectral resolution with short exposure times and a large field of view, as well as the ability to measure polarization \citep{MScavallini06}.\\
The region observed was AR11005. This region, as seen by both SOHO and Hinode images, appears as a small pore with a light bridge in the northern hemisphere at high latitude [25.2$^{\circ}$ N, 10.0$^{\circ}$ W], therefore very likely belonging to solar cycle $24$.\\
The dataset consists of $80$ sequences of measurements, each containing a $21$ point scan of of the full Stokes vector for both the $Fe~617.3~nm$ and Ca $854.2$ nm lines.
The difference in wavelength between the spectral points for the Fe line was $2~pm$. The exposure time for each image was set to $80~ms$ and each spectral scan took $52$ seconds to complete. The pixel scale of these $512 \times 512$ images was set at $0.167~arcsec$.
For each spectral image, we also acquired a broad-band (WL) and a G-Band counterpart, both imaging approximately the same FoV.
The pixel scale of the $1024 \times 1024$ WL  image ($621.3 \pm 5 nm$) was set at $0.083~arcsec$ and the exposure time was $80~ms$ (shared shutter with IBIS spectral images). The pixel scale of the $1024 \times 1024$ G-band  image ($430.5 \pm 0.5~nm$) was set at $0.051~arcsec$ and the integration time was $10~ms$. \\
The ancillary images have been restored with Multi-Frame Blind Deconvolution (MFBD) \citep{MSnoort05}, resulting in a single frame for each scan both for the G-band and the broad-band images. Using these restored images, the spectropolarimetric images have been registered and destretched to minimize the seeing effects uncorrected by the AO and to achieve the highest spatial resolution.\\
The pipeline for IBIS data reduction takes care of normal calibration processes (dark frame, flat field, etc.) and also corrects for blue-shift effects \citep{MSreardon08} and instrumental polarization: the latter is important for minimizing residual cross-talk between the Stokes profiles. For further details on the calibration pipeline see \cite{2009ApJ...700L.145V} \\
The estimated mean spatial resolution of the line-of-sight (LoS) velocity fields computed from the spectropolarimetric scans used in this work is $0.36$ arcsec.\\
In addition to IBIS data, we also used Hinode SOT/SP observations of the same active regions taken three and half hours before the IBIS data set.

\subsection{Ca 854.2 nm and Fe 617.3 nm formation heights}
When dealing with waves, knowledge of the formation heights of the spectral lines and their position with respect to the equipartition layer (namely the layer in which the MHD mode conversion takes place), is fundamental. As mentioned earlier, IBIS observations were obtained using two spectral lines: the chromospheric Ca $854.2$ nm and the photospheric Fe $617.3$ nm.\\
As for the Ca $854.2$ nm line, \cite{2008A&A...480..515C} and \cite{2009A&A...494..269V} pointed out that this low chromosphere spectral line actually spans a wide range of atmospheric heights starting from the mid-high photosphere to the low chromosphere (line core).\\
\citet{2006ASPC..358..193N} estimated the quiet Sun height of formation of the Fe $617.3$ nm core to be around $300$ km above the photosphere. This has to be compared with the altitude of the equipartition layer in our region of interest in order to estimate the plasma $\beta$ regime sampled by our observations. For this purpose, we estimated the equipartition layer position using spectropolarimetric inversions obtained through the SIR code \citep{1992ApJ...398..375R} performed on SOT/SP data. Our estimate reveals that the equipartition layer is slightly below the height of formation of the Fe $617.3$ nm spectral line throughout the FoV, this means that we are sampling the onset of the low-$\beta$ regime in the solar atmosphere, very close to the conversion altitude. 

\begin{figure*}
\centering
\subfigure[Ampl. @ $3$ mHz]{\includegraphics		[width=6cm,trim=12mm 3mm 12mm 3mm, clip]{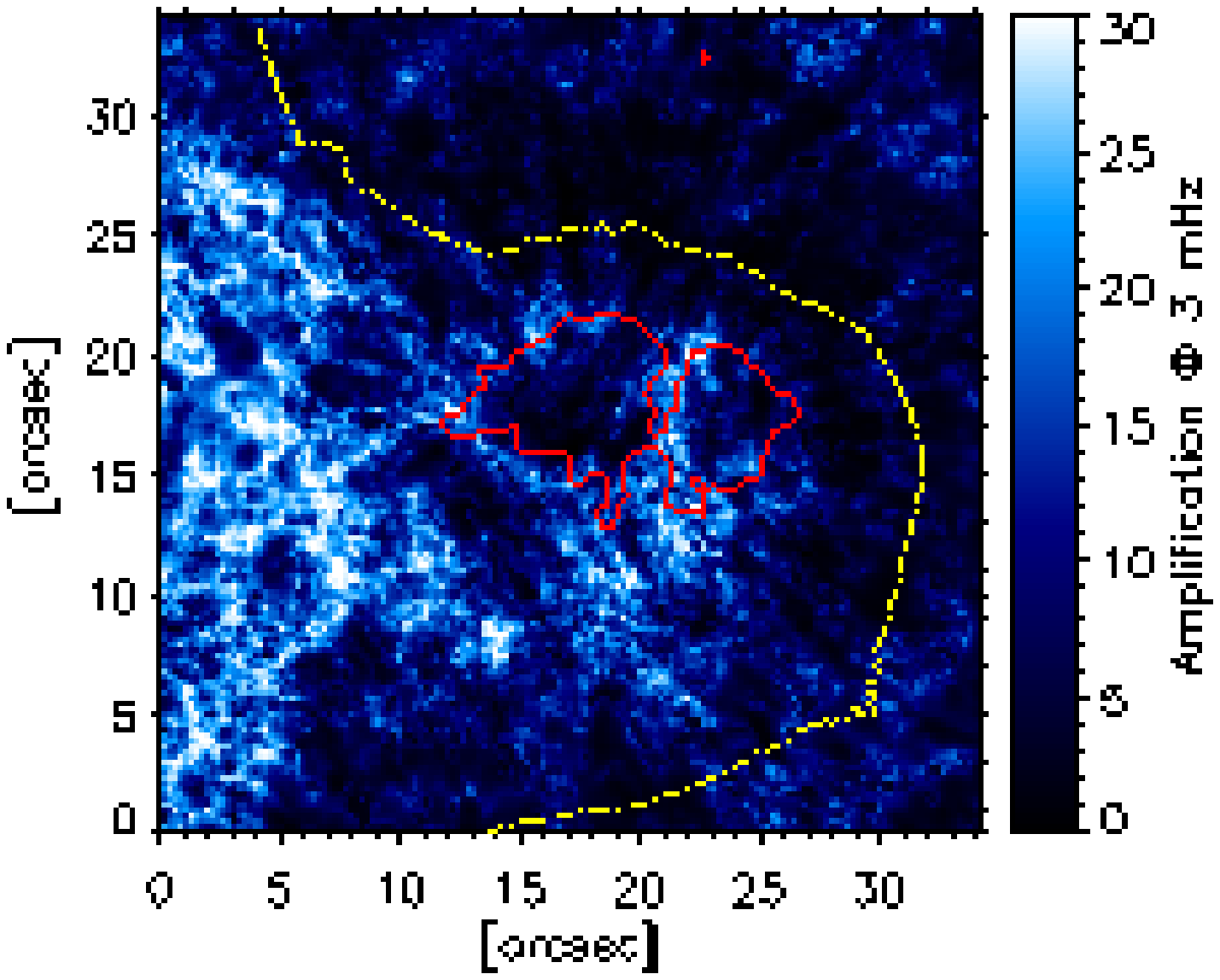}}
\subfigure[Phase lag @ $3$ mHz]{\includegraphics	[width=6cm,trim=12mm 3mm 12mm 3mm, clip]{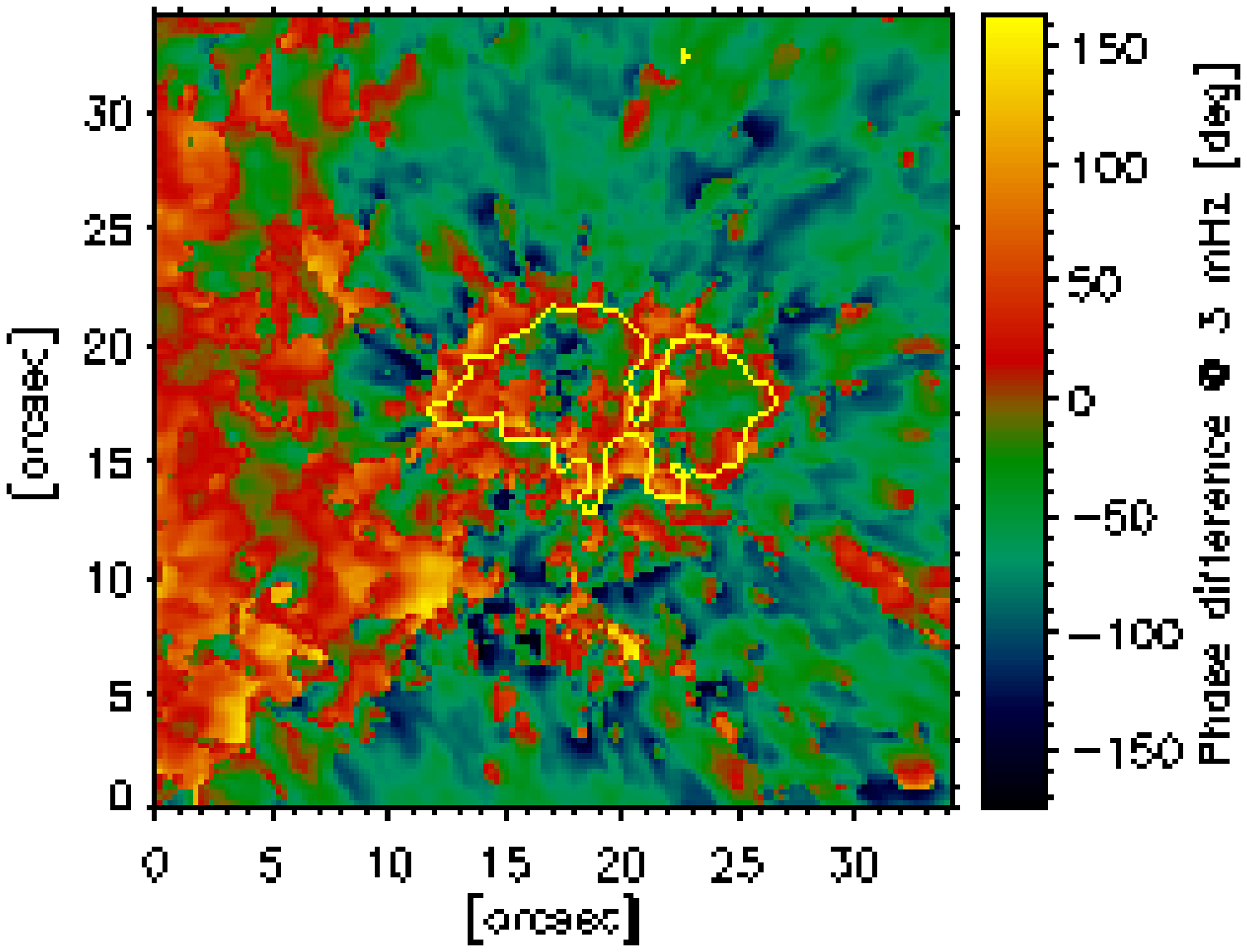}}
\subfigure[Ampl. @ $5$ mHz]{\includegraphics		[width=6cm,trim=12mm 3mm 12mm 3mm, clip]{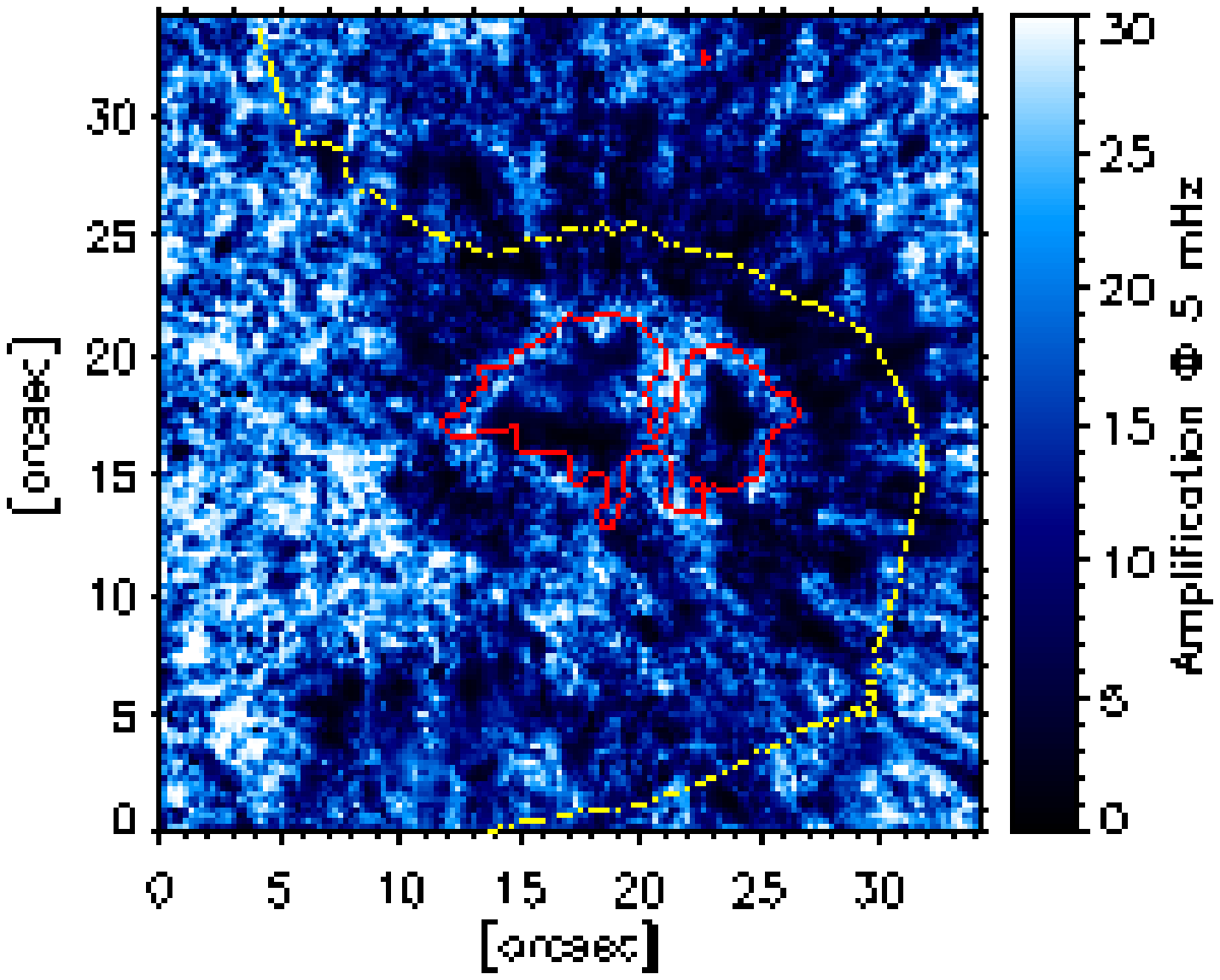}}
\subfigure[$\theta$ @ $1000$ km]{\includegraphics	[width=6cm,trim=12mm 3mm 12mm 3mm, clip]{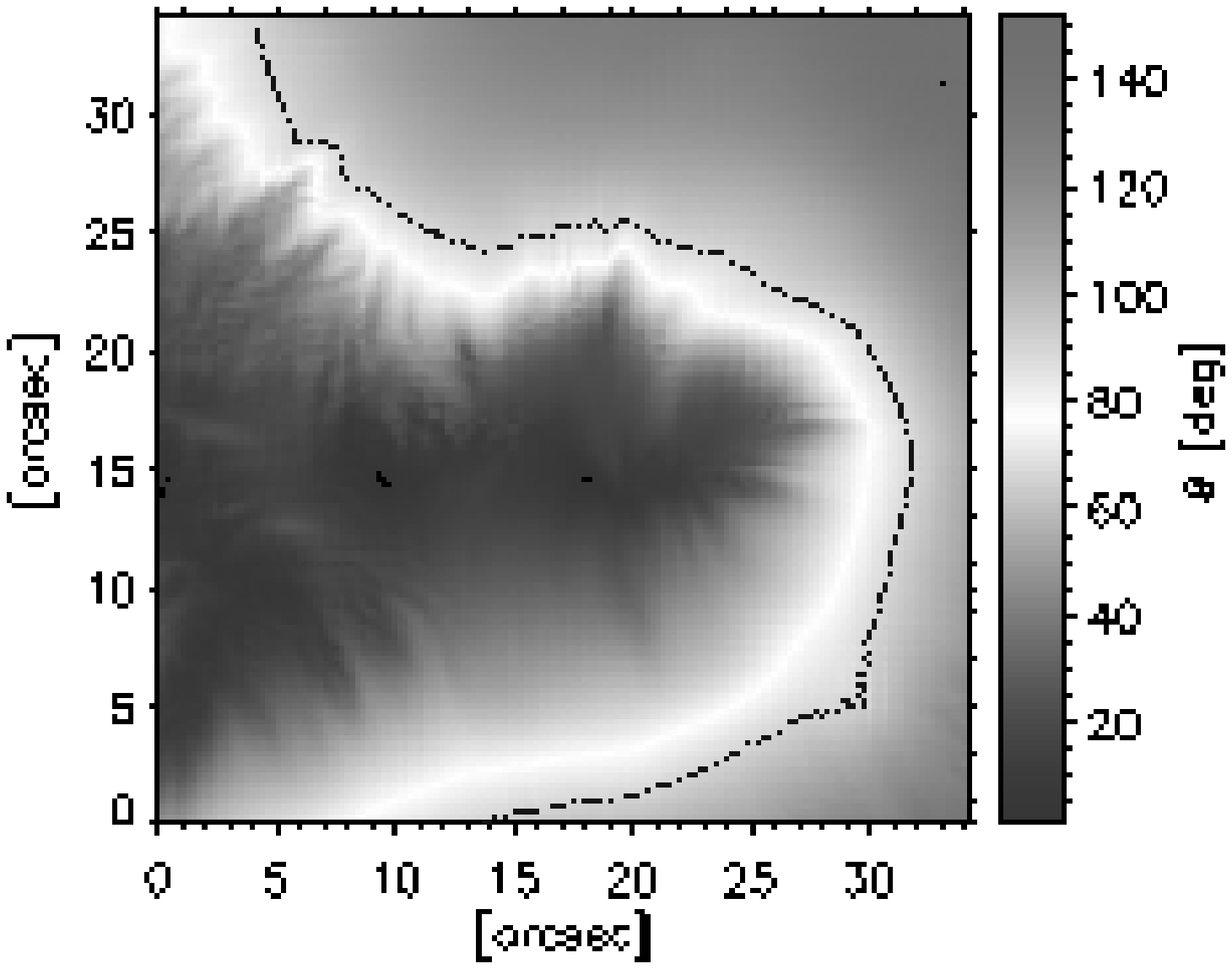}}

\caption{(a) Amplification map at $3~mHz$ from IBIS data. (b) Phase lag between photosphere and chromosphere at $3~mHz$ (positive values mean upward propagation) from IBIS data. (c) Amplification map at $5~mHz$ from IBIS data, (c) chromospheric magnetic field inclination as inferred from extrapolations of Merlin magnetic vector map of Hinode data. Continuous contours indicate approximatively the position of the umbra, dashed contours in the amplification maps indicate the position where the chromospheric magnetic field is $90$ degrees inclined with respect to the LoS.}
\label{panels}
\end{figure*}

\subsection{Inclination angle at photospheric and chromospheric heights}
As mentioned, in this work we explore the characteristics of the power spectral density at different magnetic field inclinations with respect to the local gravity vector. This is done by comparing the information encoded in the temporal sampling to the spectro-polarimetric data provided by the Hinode SOT spectropolarimeter.\\
Using MERLIN inversions we studied maps of inclination of the same region observed with IBIS a few hours later. Hinode SOT/SP Inversions were conducted at NCAR under the framework of the Community Spectro-polarimtetric Analysis Center (CSAC; \url{http://www.csac.hao.ucar.edu/}). The result is shown in fig. \ref{inclination}. In fig. \ref{panels} (panel d) we show the chromospheric magnetic field inclination obtained from non-linear force-free extrapolations performed using the NLFF code \citep{2009ApJ...696.1780D}, after resolving the azimuth ambiguity by means of the code presented in \citet{2009ASPC..415..365L} \footnote{Further information on the details of the code can be found at www.cora.nwra.com/AMBIG/}.     
Even though the low chromosphere is not a place where the force-free approximation holds, the observational results obtained in this work can easily be explained by our extrapolated model, suggesting the validity of such a scheme even in this context.

\section{Results}
In the following we will focus on the analysis of the velocity field perturbations in two power spectral bands namely: $2.8-3.8$ mHz and $4.8-5.8$ mHz, hereafter the $5$-minute and $3$-minute bands respectively. \\
Due to the limited extent of the time series data, the use of the FFT for the estimation of the power spectral density may result in a distorted estimation \citep{1970WRR.....6.1601E}. For this reason, we estimated the power spectral density using the Blackman-Tukey method and making use of a Barlett windowing function \citep{blackman1958measurement}.

\subsection{Photospheric power spectral density}
Fig.\ref{power} shows the power maps for the $3$ and $5$ minutes bands sampled by the Fe $617.3$ nm line (panels a and c respectively).\\
The power distribution in the two spectral bands look quite different. The $5$-minute oscillations tend to be spread throughout the FoV, while they are absent (absorbed) in the magnetic region. This is not surprising and a long series of similar results can be found in the literature (see \citet{2009ApJ...706..909C} and references therein). The magnetic field is, in fact, able to scatter and absorb the acoustic field thus producing such an effect. On the other hand, $3$-minute oscillations are only found in the umbra of the pore.

\subsection{Chromospheric power spectral density}
The chromospheric power maps in the $5$-minute and $3$-minute bands (panels b and d of fig. \ref{power}) are even more noticeable and interesting than their photospheric counterpart.
What is clear in these maps is the strong presence of $5$-minute oscillations in the diffused magnetic field and in the light bridge.\\
What is even more interesting is the absence of power in an annulus surrounding the spot. Our interpretation of this mechanism is that the slanted nature of the longitudinal waves running along the field lines makes them invisible to a line of sight observation. This is, of course, fitting the general view that describes the magnetic field as a filter acting on the transmitted waves and preferring those which are more aligned with the field lines themselves \citep{Cally15022006}.

\subsection{Chromospheric amplification}
In accord with the conservation of energy, the amplitude of acoustic waves in the solar atmosphere increases with height in the atmosphere due to the rapid decrease in density with increasing height. Moreover these waves are subject to non-linear steepening that leads to the formation of shocks \citep{2009A&A...494..269V}. It is therefore interesting to study the nature of this growth in amplitude as it may reveal important details about the density stratification of the atmosphere overlying the photospheric magnetic structure. It may also provide valuable information about the energy deposition brought by waves in the low/mid chromosphere.\\
In the structure under investigation, this analysis has been carried out taking the ratio of the chromospheric over photospheric power spectral density in both frequency bins. 
In the case of waves above the cut-off frequency, it is generally expected to see an upward propagation toward the chromosphere, in this sense it is not surprising to find a power amplification everywhere in the FoV as shown by the fig. \ref{panels} for the $3$-minute waves (panel c). What is equally interesting in the map is the clear evidence of a lack of amplification in the region surrounding the umbra in both spectral bands. We interpret this as evidence for the presence of slow longitudinal magneto-acoustic modes running preferentially along slanted magnetic field lines. This behavior is still present in the $5$-minute band even though it is less evident. The dashed lines in panels \textit{a} and \textit{c} of fig. \ref{panels} indicate the $90$ \degr isocontour ofthe chromospheric magnetic field inclination estimated using the field extrapolations discussed above, and corrected for the position of the region on the solar disk. As the magnetic field lines are highly inclined with respect to the LoS, the velocity field fluctuations disappear as they are longitudinal, producing an "absorption" halo in our chromospheric power and consequently in the amplification maps. It is worth noting that the absorption halo surrounding the spot and the $90$ \degr chromospheric inclination isocontours perfectly match, especially for $3$-minutes wave, although they came from independent estimates; namely power spectra analyses  and magnetic field extrapolations.\\ 
Moreover, the amplification of the $5$-minute waves is not present throughout the FoV but it is concentrated in the small magnetic elements and on the edge of the umbra. As for the small magnetic elements, this effect appears to be evidence of the lowering of the cut-off frequency due to radiative losses \citep{2008ApJ...676L..85K}, which allows the channeling of $5$-minute power in the low chromosphere even in vertical magnetic fields. This is in agreement with previous analyses of the chromospheric power spectra in small magnetic flux tubes \citep{2009ApJ...692.1211C}. On the other hand, the chromospheric amplification at the edge of the umbra can be interpreted as the lowering of the cut-off frequency due to magnetic field inclination \citep{2006ApJ...648L.151J}.

\subsection{Propagation of $5$-minute waves toward the chromosphere: phase spectra}
One of the main advantages of dual line investigations of the the solar atmosphere, is the possibility to study the vertical behavior of waves and to look for upward propagating components. For this purpose we used a phase difference analysis.\\
Fe $617.3$ nm and Ca $854.2$ nm lines are scanned sequentially in about $52$s, therefore when estimating the phase lag, using the FFT cross-correlation, we took into account the time lag between the two scans of the lines, and corrected the phase lag maps accordingly. The results of this analysis for the 5-minute band is shown in panel \textit{b} of fig. \ref{panels}. In our sign convention, positive values of the phase lag mean that the chromospheric signal is lagging behind the photospheric one, thus representing an upward propagation.\\
Similar to the amplification maps, the phase maps indicate that a lot of $5$-minute waves are reaching the formation height of the Ca $854.2$ nm.

\subsection{Power as a function of magnetic field inclination}
To investigate the properties of the magnetic field inclination in terms of wave transmission/conversion, we compare the periodogram of the observed perturbations with the photospheric inclination of the magnetic field inclination estimated by the inversions performed on the same active region observed by Hinode/SOT.\\
In particular, we chose different regions over the FoV, pertaining to different magnetic field angles to the vertical. For each of these regions we estimated the maximum of the power averaged over a $6$-degree interval, using the Welsch method, for different magnetic field inclinations.
Before performing the analysis, the Hinode and IBIS data sets were co-registered to sub-pixel accuracy using a technique based upon FFT cross-correlation.\\ 
A visual inspection of the resulting co-registration shows very good matching between the two data-sets despite being acquired 3 1/2 hours apart. This suggests that at least the strong magnetic field structures are quite stable over time, and the evolutionary time scales are much greater than the time interval between the acquisition of each data set. This further proves the feasibility of our investigation when comparing Hinode inversions and the dynamics sampled by IBIS.  
We assume that the inclination angle is equal to the attack angle between the incoming acoustic waves wave-vector and the magnetic field lines themselves. Once again we selected two representative spectral bands around $3$ and $5$ minutes. Fig. \ref{trasmissione} shows the integrated power in the two selected spectral bands as a function of magnetic field inclination.
The most striking result of this analysis is the dependence of the power upon the magnetic field geometry. This result supports the predictions of the model proposed by \citet{2006MNRAS.372..551S}.\\

\subsection{A simple model}
\begin{figure}
   \centering
   \includegraphics[width=7cm]{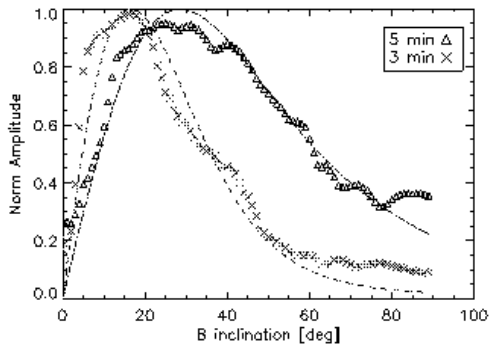}
\caption{Transmission in the $3$ and $5$ minutes bands as a function of magnetic field inclination from IBIS data. Dashed and dot-dashed curves represent the result of the fit of our model, for $3$ and $5$ minutes bands respectively.}
         \label{trasmissione}
   \end{figure}

The transmission of the $5$- and $3$-minute acoustic waves strongly depends on the attack angle which is acting as a filter.\\
To better understand this behavior we propose a simple model made up of three components related to the transmission coefficient, the ramp effect and the geometrical projection effects. Before going into the details of these terms and their contribution to our model, we introduce a modified transmission relation as follows:

\begin{equation}
T=exp \left[-\pi k h_{s} sin^{2}\theta \right]  ~ \left[1- exp \left(-\frac{\theta}{\theta _{c}} \right)\right] ~cos~ (\theta- \theta_{pos})
\end{equation}
where $\theta_{pos}$ is the heliocentric angle taking into account projection effects.
As discussed above, the first exponential term represents the transmission coefficient from fast (acoustic) to slow (acoustic). These waves are generally longitudinal waves showing a slanted propagation along the field lines. For this reason, as the magnetic field bends toward the horizontal we do not expect to find their signature, therefore we add a $cos(\theta-\theta_{pos})$ term to take into account the suppression of power due the projection effects. In doing this, we are assuming that the most important contribution is given by longitudinal acoustic waves. The middle term models the ramp effect, that is, the lowering of the cut-off frequency due to the magnetic field inclination. For inclination angles larger than a critical value $\theta _{c}$, we begin to see transmission. We then estimate the critical angle using:

\begin{equation}
\theta _{c}=cos~(\nu / \nu _{c})^{-1}
\end{equation}
where $\nu _{c}$ is the quiet Sun acoustic cutoff frequency set to $5.2$ mHz.\\
Fig. \ref{trasmissione} shows the normalized transmission amplitude as a function of inclination angle superimposed with the trend expected from the model (dashed lines). Interestingly, using the model itself, we can estimate the width of the layer in which the conversion is taking place (the $h_{s}$ parameter). It appears that this parameter is different for the $5$-minute and $3$-minute waves.\\
From a least-squares fit of our data, we found $h_{s}=242$ km for the $3$-minute waves and $203$ km for the $5$-minute waves, with a reduced $\chi^{2}=0.0068$ and $\chi^{2}=0.0052$ respectively. \\
The reason for this difference is found in the different spatial location of the two kinds of waves. As shown in fig. \ref{panels} and commented above, $3$-minute waves and $5$-minute waves are not co-spatial, the former are mainly located inside the umbra of the pore while the latter are located in the surrounding region and in the diffused magnetic field. The conversion layer thickness depends on the slope of the $c_{s}^2/a^2$ ratio and hence it depends on the local atmospheric conditions. Our estimates of the $h_{s}$ parameter suggest that in the umbral region, where the power of the $3$-minute waves manifests, the slope of $c_{s}^2/a^2$ is smaller than that associated to the $5$-minute waves and thus the conversion layer for the $3$-minute waves is thicker.

\begin{figure}[t]
\centering
\subfigure{\includegraphics[width=7cm]{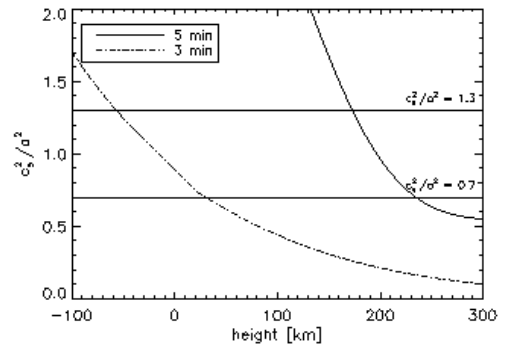}}
\caption{Estimation of the ratio $c^2_{s}/a^2$ in the same regions selected for the analysis of the transmission obtained through SIR spectropolarimetric inversions of Hinode data. The region emitting $5$ minutes waves show a steeper $c^2_{s}/a^2$ ratio, suggesting a smaller conversion layer thickness.}   
\label{beta} 
\end{figure}  

To independently confirm that, we estimate the shape of the $c_{s}^2/a^2$ function in the same regions selected for the analysis of the transmission, by using SIR spectropolarimetric inversion code \citep{1992ApJ...398..375R} to produce an atmospheric model starting from the Hinode data set. The results of this analysis is reported in fig. \ref{beta} where the above mechanism is proved.\\
The regions emitting $5$-minute waves show a steeper $c_{s}^2/a^2$, thus suggesting a smaller conversion layer thickness with respect to the regions emitting $3$-minute waves (the umbral region), where the thickness is larger. 

\section{Discussion}
The results presented so far point to a fairly self-consistent picture in which the amount of energy provided by waves in the upper Sun's atmosphere strongly depends on the magnetic field topology. Our results also suggest that most of the waves transmitted in the chromosphere are longitudinal waves, that is, slow (acoustic-like) MHD waves in low-$\beta$ atmosphere, whose signature disappear from our LoS as the magnetic field lines become highly inclined. We believe that this interpretation is supported by the maps of the chromospheric power and the amplification showing a clear absorption in those regions where the magnetic field is inclined by ~$90$ \degr with respect to the LoS.\\
With this in mind we have been able to modify the transmission coefficient for acoustic waves crossing the equipartition layer in the solar atmosphere to provide a simple model for our results.\\
Our model is able to fairly well reproduce the observed dependence of the power to the magnetic field photospheric inclination angle for both $5$ minutes and $3$ minutes waves once the local modification of the cut-off frequency is taken into account. This  provides us with an estimate of the thickness of the equipartition layer in which the mode conversion takes place ($h_{s}=200-240$ km).\\
As highlighted by \citet{2006MNRAS.372..551S}, the magnetic field causes the helioseismic waves to split into fast and slow magnetoacoustic branches and it causes the waves to be progressively more field-aligned. Our results suggest that in the chromosphere, most of the energy contribution due to waves is in the slow acoustic mode, that is, longitudinal and field aligned waves.\\
Moreover we also find a small discrepancy in the estimate of the equipartition layer thickness which can be explained by the different slopes of the $c_{s}^2/a^2$ as a function of height corresponding to the selected regions, as inferred from spectropolarimetric inversions. We believe that this provides a further test of the validity of our model. Since the $5$-minute waves and the $3$-minute waves are not co-spatial, they reflect the changes of the atmosphere in which they are running.

\section{Conclusions}
In this work we addressed the problem of wave transmission in different magnetic field inclination regimes in the solar atmosphere. This problem has been investigated by means of IBIS multi-height observations and by modeling them with a general theoretical picture based upon fast to slow transmission theory.\\
Our results clearly underline the presence of a strong connection between wave transmission and magnetic field geometry. Using spectropolarimetric IBIS data, we studied the behavior of the photospheric power in two spectral bands, namely $5$ and $3$ minutes, with the inclination angle of the observed magnetic field, as inferred from Merlin inversions of Hinode data. We found that the power is not equally distributed but peaks at certain angles of inclination: around $25\degr$ for $5$ minutes and $15 \degr$ for $3$ minutes waves. \\
Using a basic model that includes a fast to slow transmission term, projection effects expected in the presence of longitudinal waves, and the so-called ramp effect, we have been able to reproduce the observational signature of the power as a function of the inclination angle. This has revealed that most of the waves observed in the photosphere are longitudinal acoustic waves characterized by a slanted propagation along the field lines.\\
This scenario is also supported by the power spectrum analysis at both photospheric and chromospheric layers, which reveals an evident lack of power and of amplitude amplification at regions where the field lines are expected to be inclined of $90\degr$ to the LoS. This emphasizes that the lack of waves signature is due to the presence of longitudinal waves (slow MHD mode in low-$\beta$ chromospheric plasma).\\
We mainly found upward propagating waves in both $5$ and $3$ minutes able to reach to chromosphere. This happens predominantly at the edges of the magnetic umbra and in the diffused magnetic field surrounding.

\begin{acknowledgements}
We acknowledge Fabio Giannattasio for the SIR inversions of Hinode data. We also acknowledge the HAO-CSAC team for the Merlin inversion client. IBIS was built by INAF-Osservatorio Astrofisico di Arcetri with contributions from the Università di Firenze and the Università di Roma "Tor Vergata". Hinode is a Japanese mission developed and launched by ISAS/JAXA, collaborating with NAOJ as a domestic partner, NASA and STFC (UK) as international partners. Scientific operation of the Hinode mission is conducted by the Hinode science team organized at ISAS/JAXA. This team mainly consists of scientists from institutes in the partner countries. Support for the post-launch operation is provided by JAXA and NAOJ (Japan), STFC (U.K.), NASA, ESA, and NSC (Norway). 
\end{acknowledgements}

\bibliography{stangalini}
\bibliographystyle{aa}
\end{document}